\documentclass[12pt,A4]{article}
\input{epsf.tex}
\begin{document}
\begin{center}
{\bf PARAMETERS OF DISTRIBUTION OF THE PRIMARY GAMMA-TRANSITION INTENSITIES
FOLLOWING RESONANCE NEUTRON CAPTURE AND SOME PROPERTIES OF COMPOUND NUCLEI
$^{157,159}$Gd}\\\end{center}\begin{center}
{\bf A.M. Sukhovoj, V.A. Khitrov}\end{center}\begin{center}
{\it Joint Institute
for Nuclear Research, 141980, Dubna, Russia}
\end{center}
\begin{abstract}
The re-analysis of the published experimental data on the primary
gamma-transition intensities following neutron capture in different
groups of neutron resonances in $^{156,158}$Gd has been performed.
There are determined the most probable values of sum of E1 and M1
primary transitions, numbers of excited by them levels of both parities,
ratios of radiative strength functions  k(M1)/k(E1),
dispersions of deviations of random values of intensities from the
average and ratios of mean intensities of primary transitions
to levels J=5/2 with respect to analogous data for J=1/2 and 3/2
(capture of the 24 keV neutrons) in narrow excitation energy intervals.

All the data on level density and sums of radiative strength functions
confirm the presence of clearly expressed step-like structure in level
density below 3 MeV and general trend in change in strength functions
as changing primary gamma-transition energy.
Variations of distribution dispersions and, especially, ratio k(M1)/k(E1)
(or $k(E1)/k(M1)$) at changing excitation energy point to strong change
in structure of these nuclei above 1.0-1.5 MeV.

As in neighboring isotopes $^{156,158}$Gd, the shape of energy dependence
of $k(M1)+k(E1)$ considerably differs as  changing nuclear mass.
This can be due to dependence of gamma-decay process on structure of
neutron resonance and/or levels excited by gamma-transitions.
The dilemma appeared can be solved only in direct experimental search
for structure of neutron resonances in region of their energy of about
two nucleon pairing energy in nuclei of corresponding mass.
\end{abstract}

\section{Introduction}\hspace*{16pt}\hspace*{16pt}

Density of excited levels $\rho$ and emission probability $\Gamma$ of any
nuclear reaction product following neutron resonance $\lambda$ decay are
the main sources of experimental information on properties of nuclear
matter and practically important nuclear-physics constants.
This stipulates for necessity of their determination with maximum possible
accuracy.

However, the $\rho$ and $\Gamma$ values cannot be determined for main mass of
nuclei excited in $(n,\gamma)$ reaction in direct experiments: mean spacing
$D$
between levels is comparable with or much less than a resolution (FWHM) of
existing spectrometers. Correspondingly, these parameters can be extracted
only from the spectra measured with ``bad" resolution.
Only one-step reactions were mainly used for this aim.
Obtaining of information on $\rho$ and $\Gamma$ from gamma-spectra of  two-step
reaction was started only in the last time \cite{Meth1}.
Comparison of shape of functional dependence of the obtained in this way
values of $\rho=f(E)$ and radiative strength functions
$k=\Gamma_\lambda/(E_\gamma^3 D_\lambda A^{2/3})=\phi(E_\gamma)$ with the data
of one-step reactions (spectra of evaporated nucleons, different gamma-spectra)
points to their principle incompatibility.
It appears itself in presence \cite{Meth1} or lack of abrupt changes in
determined parameters. This operation allows one to determine sources of
systematical errors, estimate their values and reveal the region of maximum
discrepancy between the data of different experiments.

Analysis of the most sufficient sources of systematical errors \cite{TSC-err}
and their transfer coefficients onto errors of parameters in practically
realized case of two-step reaction (two simultaneously emitted gamma-quanta)
showed that even maximum possible errors of the $\rho$ and $\Gamma$ values
(obtained from two-step reaction)
cannot explain discrepancy for one- and two-step reactions.

Nevertheless, necessity of additional testing a method for determination of
$\rho$ and $\Gamma$ from two-step reactions calls no doubts.
At present, such test can be performed only in model-less analysis of the
primary gamma-transition intensities from reaction $(\overline{n},\gamma)$
(capture in ``averaged"~resonances).

Possibility to obtain new information from these data is caused by the use
by authors of experiments of unnecessary for analysis ideas of ``statistical"~
mechanism of gamma-decay process and abstract form of distribution law of
arbitrary gamma-transition intensity deviation in individual resonance from
mean value. The shape of approximating function and concrete results of
analysis of rather limited set of strongly fluctuating data are given in
\cite{Appr-TSC}. More general variant of analysis is described in \cite{Yb174}.

\section{Experimental data}\hspace*{16pt}\hspace*{16pt}

In the isotopes under consideration $^{157,159}$Gd were measured the
primary transition intensities following resonance neutron capture with mean
energies of 2 and 24 keV; there are also data for $^{159}$Gd for both capture
of neutrons in resolved resonances  and on beam with cadmium filter.
Experimental data cover maximum level excitation energy diapason for
even-odd deformed nuclei and allow one to get maximum possible and completely
independent information on the desired gamma-decay parameters.

In two last cases s-neutrons are mainly captured, but capture of
p-neutrons must be taken into account at $E_n$=2 and 24 keV.
According to \cite{Reich}, ratio between available in that time values of
strength functions of s- and p-neutrons does not exceed 0.08 for $E_ n$=2,
but is grater than 1 for 24 keV.
Therefore, the following analysis should account for excitation of resonances
with spins 1/2, 3/2 and different parity only for the data $E_n$=24 keV.
This means that the averaged intensity of the primary dipole gamma-transitions
to levels  5/2 depends on ratio between strength functions $S_0$, $S_1$ and
that it must be varied at determination \cite{Yb174} of distribution
parameters of the random intensity deviations from mean value.
In principle, level density with spin 5/2 in deformed nuclei must be bigger
than corresponding sum for spins 1/2, 3/2 (as it approximately follows from
known functional dependences $\rho=f(J)$).
In case of lower neutron energy, there was adopted hypothesis of equal
intensity of primary gamma-transitions with near energy and the same
multipolarity to levels with spins 1/2, 3/2.

This assumption can be mistaken for strongly differing structure of wave
functions of excited levels with different spin.
According to modern microscopic nuclear models, amplitude of gamma-transition
is determined by a set of quasiparticle and phonon components.
Their contribution depends on wave functions of both decaying and excited
levels (simplified expression for matrix element of gamma-transition in
even-odd nucleus is given for example, in \cite{Malov}).

\section{Required parameters of analysis}\hspace*{16pt}
Examples of random intensity distributions in integral form for their different
$N_\gamma$ values in given level excitation energy interval,
ratios $k(M1)/k(E1)$
of strength functions of E1- and M1-transitions, dispersions of distributions
$\sigma^2=2/\nu$ and registration threshold of peak in spectrum are given
in \cite{Yb174} for the case of excitation of resonances with the only spin
value.

Introduction of additional parameter (ratio between mean intensities of
primary gamma-transitions to levels 5/2 and corresponding data for levels
1/2, 3/2) in the case of resonance with two spin values weekly influences
stability of approximation and sensitivity of this process to variation of
initial values of fitted parameters.
This result was tested for the case when mean intensities in two different
groups differed, practically, by a factor 2.

\section{Results of analysis}\hspace*{16pt}\hspace*{16pt}

Experimental cumulative sums of relative intensities $<I_\gamma/ E_\gamma^3>$
are presented in figs. 1,2 together with their best approximations.
As an example, there are used the data for $E_n=2$ keV. Due to negligible
contribution of p-neutron capture, approximation takes into account only two
distributions -- E1- and M1-transitions following decay of resonances with
spins 1/2. Experimental distributions for $E_n=24$ keV are superposition of
four distributions -- two distributions mentioned above and one more pair
corresponding to gamma-transitions between resonances with $J$=3/2 and
final levels with $J$=5/2. Its mean intensity relatively to gamma-transition
intensities to levels $J$=1/2, 3/2 is found equal $\approx 0.44$ for both isotopes
(but with some larger dispersion of values for $^{159}$Gd).

\begin{figure}[htbp]
\vspace{4cm}
\leavevmode
%\hspace{-.8cm}
\epsfxsize=17cm

\epsfbox{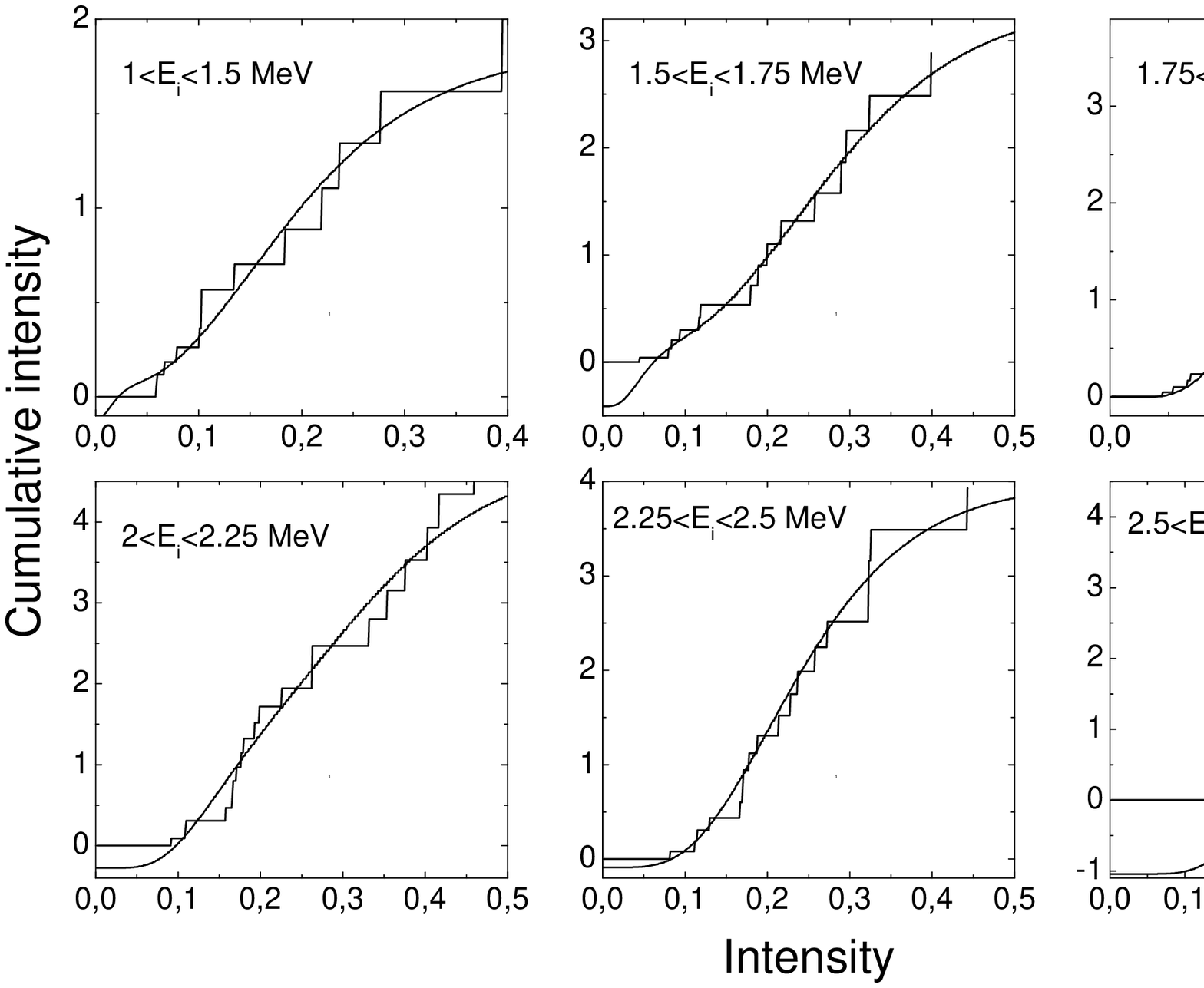}
\vspace{-7cm}

{\bf Fig.~1.}  The experimental cumulative sum of reduced intensities
$<I_\gamma/ E_\gamma^3>$ for $^{157}$Gd - histogram.
Smooth curve corresponds to the best approximation.
Excitation energy intervals of cascade final levels $E_i$ are given in figure.
Experimental data for neutron energy $\approx 2$ keV.
\end{figure}

\begin{figure}[htbp]
\vspace{3cm}

\leavevmode
\epsfxsize=17cm

\epsfbox{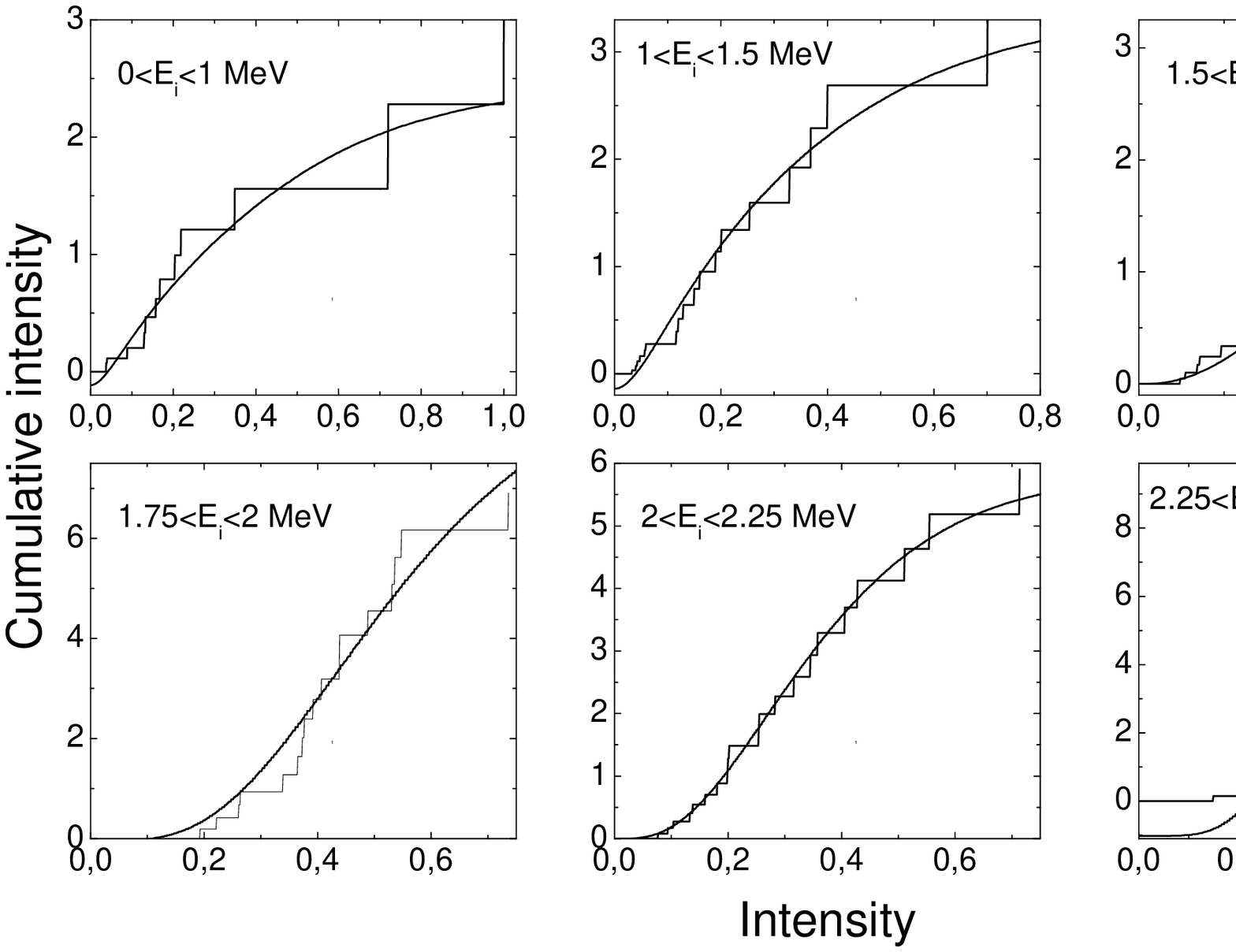}
\vspace{-7cm}

{\bf Fig.~2.} The same, as in Fig. 1, for $^{159}$Gd.
\end{figure}

All the data are presented so that the expected total intensity of
gamma-transitions lying below registration threshold  corresponds to the
most probable value of cumulative sum for zero intensity.

Precision in determination of parameters of approximating curve at low energy
of final levels $E_i$ must get worse due to inequality of level densities
with different parity. Most probably, this increases error of extrapolation by
approximating curve to zero intensity of gamma-transition.
In practice, overestimation of the $N_\gamma$ values seems to be more probable.

The best values of fitting parameters $R_k=k(M1)/k(E1)$ and $\nu$ are presented
in figs. 3 and 4. Main part of data in fig. 3  corresponds to the
case $k(M1)/k(E1)$. In potentially possible case $k(E1) \leq k(M1)$,
some portion of the data shown in these figures corresponds to alternatively
determined ratio $k(E1)/k(M1)$. These cases cannot be revealed without the
use of additional experimental information. Noticeable change in
$R_k=k(M1)/k(E1)$ and $\nu$ at $\approx 1$ MeV points to considerable change
in  structure of isotopes under study at this excitation energy.
As it is seen from figure 4, fluctuations of random intensities
to final levels of $^{159}$Gd  with spins 1/2, 3/2, from the one hand, and 5/2,
from the other hand, are described by distributions with rather different
values of $\nu$. This fact has, in principle, the following interpretation:

\begin{figure}[htbp]
\vspace{4cm}

\leavevmode
\epsfxsize=17cm

\epsfbox{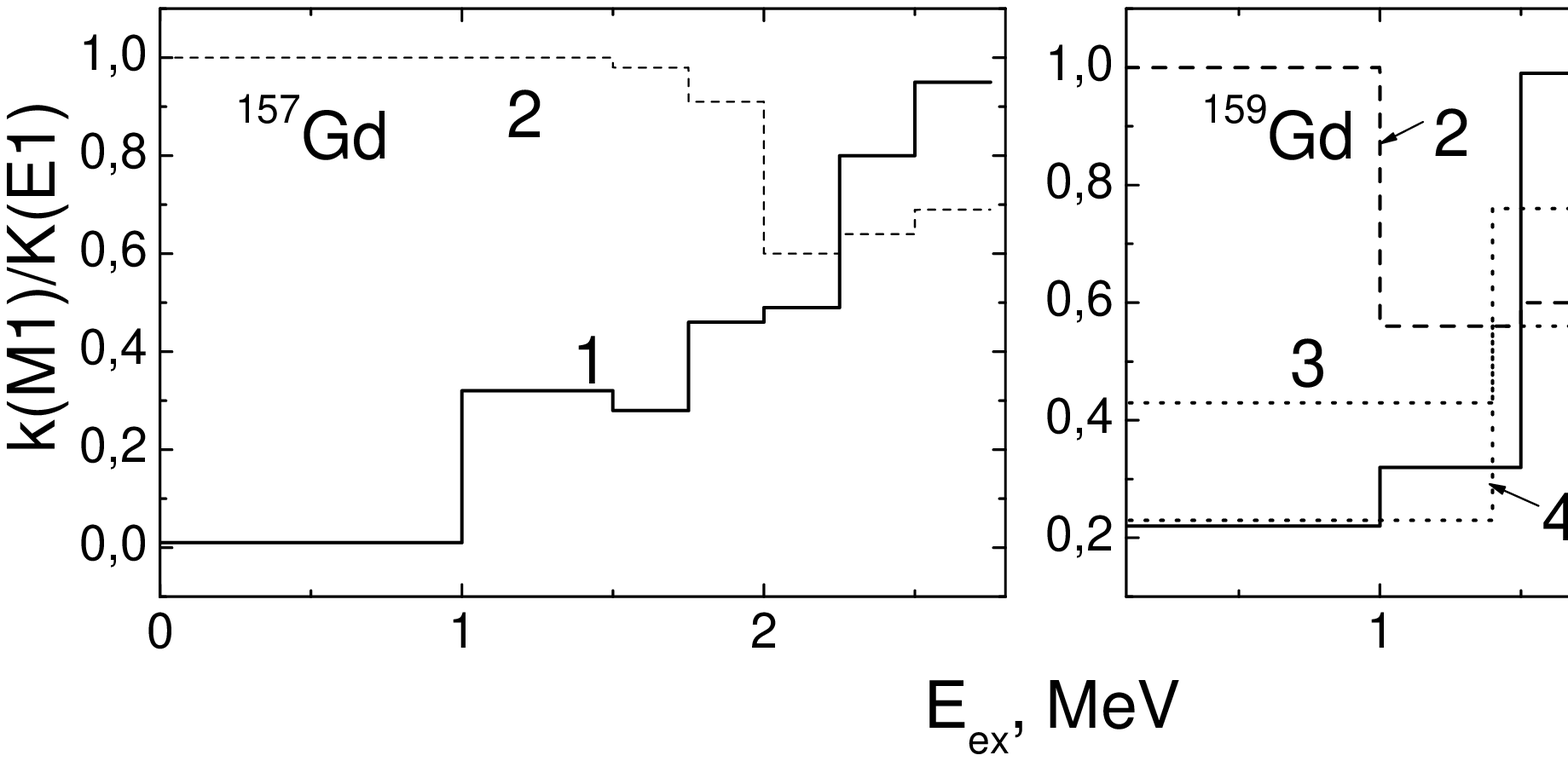}
\vspace{-10.05cm}

{\bf Fig.~3.} The best values of ratios $k(M1)/k(E1)$
(or $k(E1)/k(M1)$) for different energy of levels excited by dipole
gamma-transitions in $^{157,159}$Gd.
Line 1 represents data for $E_n \approx 2$ keV, line 2 - for
$E_n \approx 24$ keV. Line 3 -- data for
$E_n >1$ eV, line 4 -- data for isolated resonances
\end{figure}

\begin{figure}%[htbp]
\vspace{4cm}

\leavevmode
\epsfxsize=17cm

\epsfbox{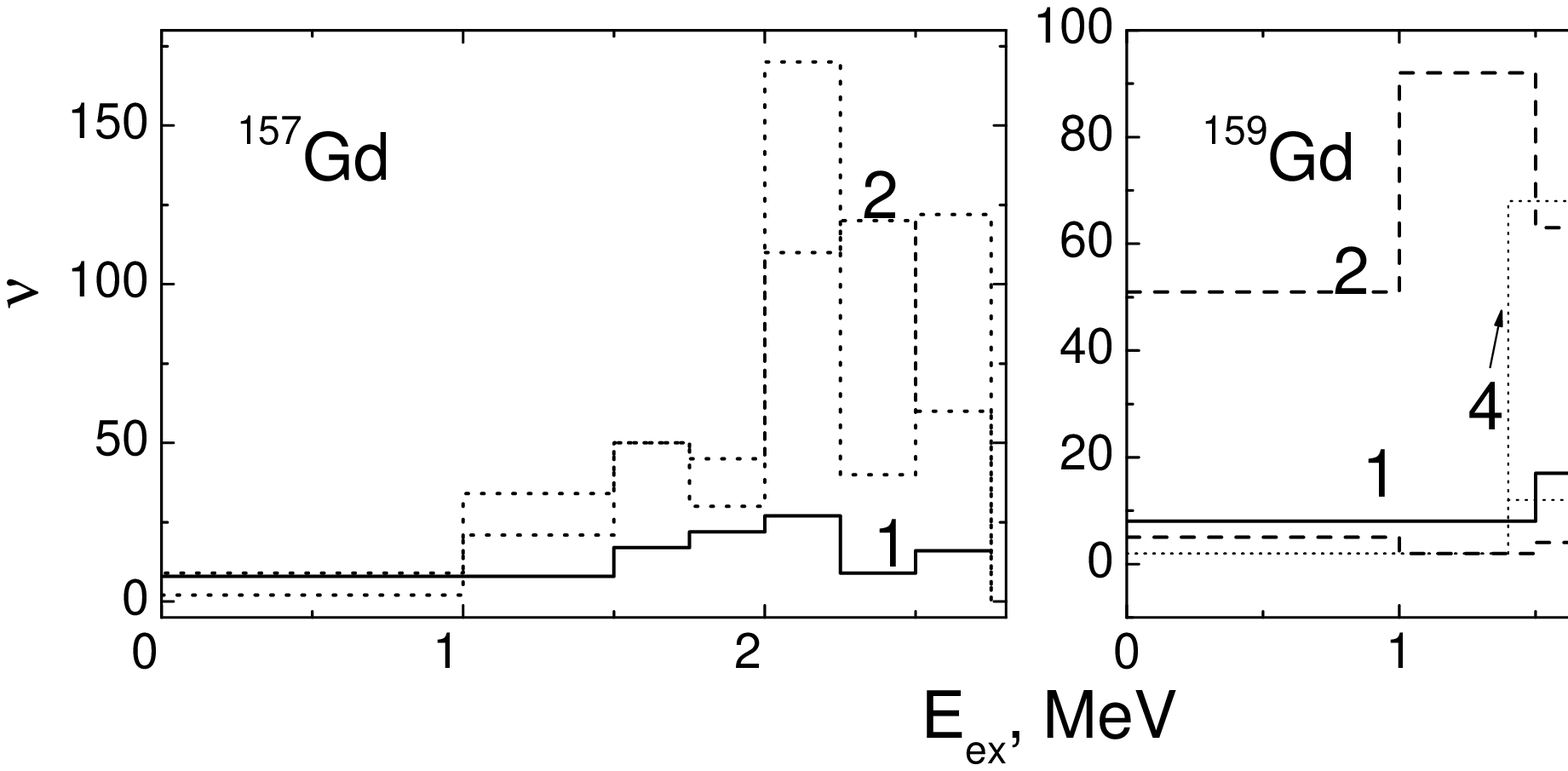}
\vspace{-10.05cm}

{\bf Fig.~4.} The same, as in Fig. 3, for parameter $\nu$ for $^{157,159}$Gd.

\end{figure}
(a) distinction in structure of decaying neutron resonances with different
spin and parity;

(b) amplitude of corresponding gamma-transitions is determined by components
of wave functions which differ  (see, for example, \cite{MalSol}) in number
of phonons and degree of their fragmentation;

(c) different structure and degree of fragmentation of the levels excited by
primary gamma-transitions.

In principle, one cannot exclude possibility of presence of some specific
systematical uncertainty which explains this effect.
But, there is required realistic explanation for its selectivity with respect
to resonance spins. Unreality of existence of assumed uncertainty brings
to considerable conclusion: there are not grounds to exclude possible
appearance of the different $\nu$ values for gamma-transitions with different
multipolarity to final levels with different spin.
The data in figs. 3,4 unambiguously point to evident change in structure of
wave functions of final levels below and above $\approx 1$ MeV.

\begin{figure}[htbp]
\vspace{4.cm}

\leavevmode
\epsfxsize=17cm

\epsfbox{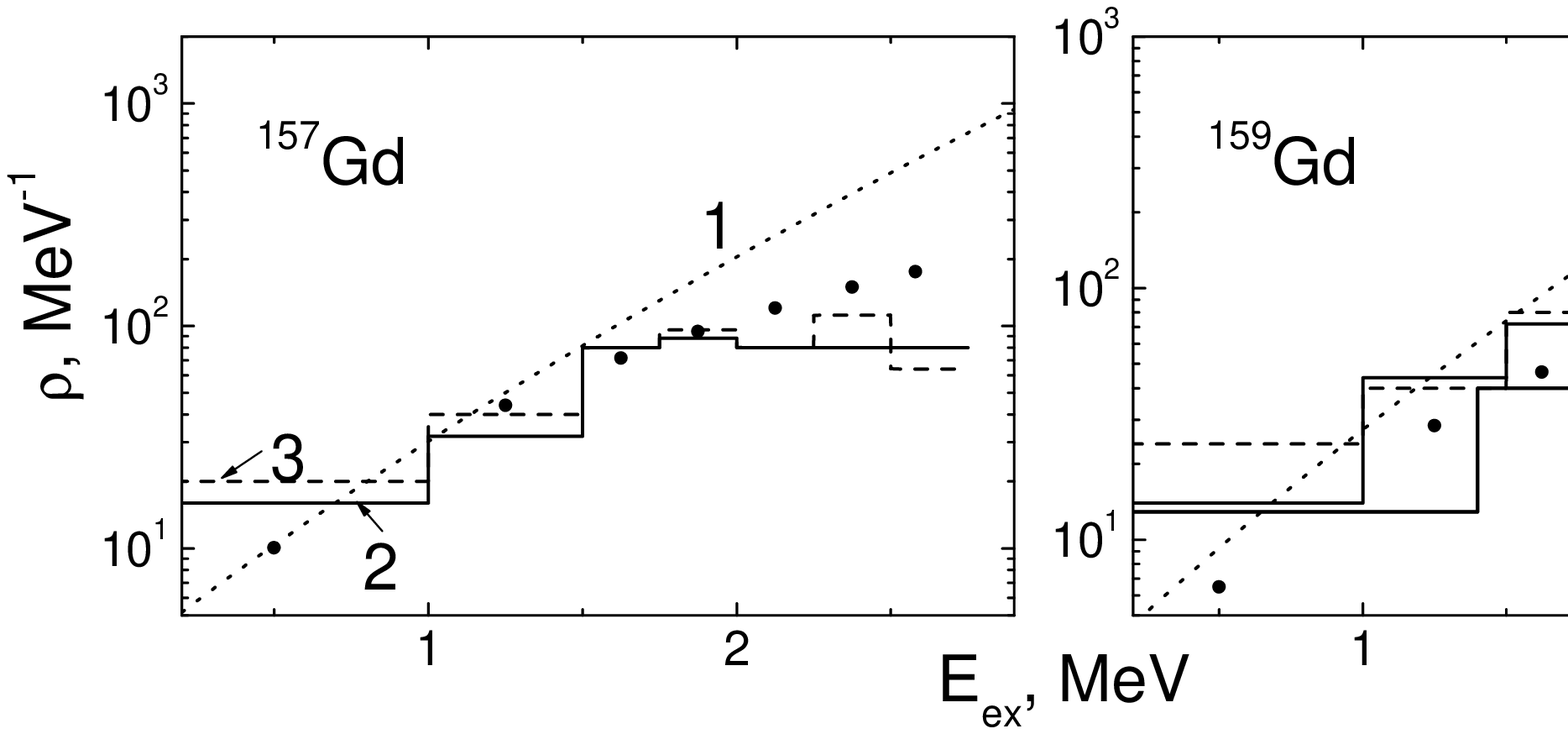}
\vspace{-10.05cm}

{\bf Fig. 5.} Different data on level density in $^{157,159}$Gd. Dotted curves 1
represent results of calculation within model \cite{Dilg}.
From the data for  $E_n \approx 2$ -- histogram 2; $E_n \approx 24$ keV --
histogram 3. Points - the best fit of data for $E_n \approx 2$ within model
\cite{Strut} for $K_{coll}$=const.
 Histogram 4 shows data
for  $E_n >1$ eV, histogram 5 - for isolated resonances.
\end{figure}

The best values of level density $\rho = \sum_{J,\pi}N_\gamma/\Delta E$ and
sums of radiative strength functions $\sum <I_\gamma>/ (E_\gamma^3 N_\gamma)$
are shown in figs. 5, 6. Normalization of intensities and strength functions in
both \cite{Gd157} and \cite{Gd159} was performed for five (from six) data sets.

\begin{figure}[htbp]
\vspace{4cm}

\leavevmode
\epsfxsize=17cm

\epsfbox{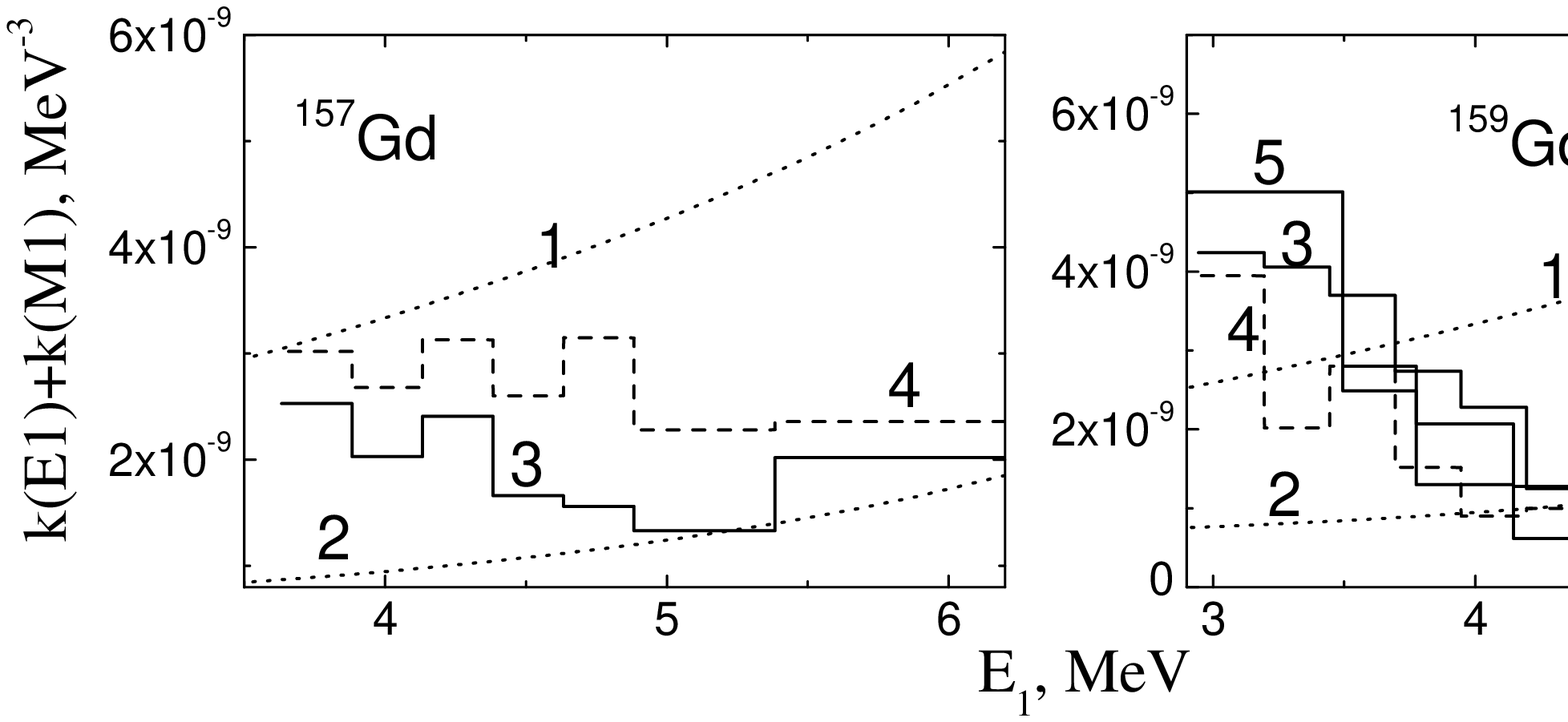}
\vspace{-10.05cm}

{\bf Fig.~6.}  Different data on radiative strength functions of the
primary gamma-transitions in  $^{157,159}$Gd. Dotted curve 1 shows results
of calculation according to model  \cite{Axel}; dotted curve 2 -- calculation
within model  \cite{KMF} in sum with $k(M1)=const$. From the data for
$E_n \approx 2$ -- histogram 3; $E_n \approx 24$ keV -- histogram 4.
From the data for  $E_n >1$ eV -- histogram 5; isolated resonances  --
histogram 6.
\end{figure}

Because intensities of gamma-transitions following ``filtered" neutron capture
are given in \cite{Gd157,Gd159} in relative units then corresponding  strength
functions in figs. 5, 6 are combined with each other under assumption of
their approximate equality for final levels $E_i <1$ MeV.

The most probable approximated $N_\gamma$ values from capture spectra of 2 keV
and 24 keV neutrons in $^{158}$Gd below 1.0 and above 2.5 MeV exceed 
``resonance"~ values, practically, by a factor of 2
(and some less - in other excitation energy intervals).

\subsection{Some sources of systematical errors}\hspace*{16pt}

Absolute minimum of $\chi^2$ for all the used sets of intensities is achieved,
practically, for the only value of $N_\gamma$, if the value of this variable
equals $\sim 5$. Change in this parameter by $\pm 1$ brings usually to
considerable increase in $\chi^2$.

This allows one not to account for possibility of considerable
(for example, more than 10-20\%) uncertainty in determined level density
caused by chosen conditions of approximation (excluded inequality of level
densities with different parity). 

Main problems in determination of nuclear parameters and their systematical
errors are, most probably, caused by:

(a) the use of assumptions on shape of the random intensity deviations from
mean value and

(b) possible presence of significant systematical errors in sets of analyzed
intensities \cite{Gd157,Gd159}.

1. The Porter-Thomas distribution allows very significant random
partial widths. But fluctuations of the measured gamma-transition intensities
$I=\Gamma_{\lambda i}/ \Gamma_\lambda$ are always limited in
their maximum value by
positive correlation between partial and total radiative widths of decaying
level. This results in some overestimation of number of degrees of freedom
$\nu$ determined by approximation and its dependence on intensities included
in approximation of cumulative sums. For instance, there is region of values
$I_\gamma/(E_\gamma^3)>0.4$ for $1<E_i <1.5$ MeV shown in Fig. 1.
Intensities in the performed analysis were normalized so that their maximum
value did not exceed $\sim 50\%$ of approximation region for the majority of
data sets.

2. The main error of analysis can be related only to ``loss" ~of gamma-transitions
whose intensities do not exceed threshold value and/or  mistaken identification
of quanta ordering in gamma-cascades.

Probability of overlap of two peaks corresponding to near-lying levels was
estimated in \cite{Stel}. As it follows from the data presented by authors,
this effect is small enough and, most probably, cannot explain significant
(several times) discrepancy between level density determined by us and its
prediction in the frameworks of the Fermi-gas model. Moreover, this overlapping
in the chosen presentation of experimental data increases rate of growth
of cumulative sums and, most probably, overestimates the $N_\gamma$ value
obtained.
There could be essential uncertainty caused by even and considerable loss of
some part of observed peaks corresponding to intense primary transitions due
to groping of levels in near-lying (spacing of about 1-2 keV) multiplets.
But this possibility is not predicted by modern nuclear theory.
Experimental data of nuclear spectroscopy also do not point to existence of
numerous multiplets of neighboring  levels with spins 1/2 and 3/2 in even-odd
compound nuclei.

3. It is also possible that the gamma-transitions in the all or greater part
of chosen intervals of of primary transition intensities (with the width of
some hundreds keV) have different mean values. Moreover, probability of
relatively low-intensity gamma-transitions quickly (but smoothly) increases
as decreasing their intensity. In principle, this effect can be caused by
mechanism of fragmentation \cite{MalSol} of different states over neighboring
levels of a nucleus.

Apparently, only this hypothesis can be alternative explanation of ``step-like"~
structure in level density in performed here analysis.
This hypothesis can be applied to the determined according to
\cite{Meth1,PEPAN-2005} level densities under the following conditions:
main part of levels (with the same $J^{\pi}$) below $\sim 0.5B_n$ must not
be excited by primary gamma-transitions;
some Cooper pairs of nucleons must break simultaneously at small as
compared with $B_n$ nuclear excitation energy. One cannot suggest other
possibility for precise calculation of the two-step gamma-cascade intensities
in the investigated even-odd nuclei.

\subsection{Interpretation of the obtained results}\hspace*{16pt}

The most important physics information on structure of excited levels below
$\approx 0.5B_n$ can be derived from coefficient of collective enhancement
of level density $K_{\rm coll}$:

\begin{equation} \rho(U,J,\pi)= \rho_{\rm qp}(U,J,\pi) K_{\rm coll}(U,J,\pi).
\end{equation}

In accordance with modern notions, $K_{\rm coll}$ determines \cite{RIPL}
degree of increase in density of pure quasiparticle excitations
$\rho_{qp}(U,J,\pi)$
in deformed nucleus due to its vibrations and rotation. One can accept in the
first approach that, to a precision of small constant, it equals coefficient
of vibrational enhancement of level density $K_{\rm vibr}$.

On the whole, this coefficient is determined by change in entropy $\delta S$
of a nucleus and redistribution of nuclear excitation energy $\delta U$
between quasiparticles and phonons at nuclear temperature $T$:

\begin{equation}
K_{\rm vibr}=exp(\delta S-\delta U/T).
\end{equation}

Now there is a possibility for unambiguous experimental determination
\cite{Prep196}  of breaking threshold $E_N$ for the first and following Cooper
pairs, value and shape of correlation functions $\delta_N$  of nucleon pair
number $N$ in heated nuclei. The main uncertainty of $E_N$ is caused by the
lack of experimental data on function $\delta_N=f(U)$, the secondary - by
uncertainty of one-quasiparticle level density $g$ in model \cite{Strut}.
So, three different model dependent approximations of level density in large
set of nuclei (\cite{Prep196} and \cite{PEPAN-2006}) predict threshold $E_2$
for five-quasiparticle excitations which differs by a factor of 1.5-2.0.

In practice, we used the second variant of notions of the Cooper pair
correlation function in heated nucleus \cite{Prep196} for estimation of the
$K_{\rm vibr}$ value from approximation of the data \cite{Gd157,Gd159}.
The values $\delta_1=1.02$ MeV, $g=9.95$ MeV$^{-1}$ were used in calculation.
Densities of three-quasiparticle levels
(multiplied by the ``best"~ $K_{\rm vibr}=const$ value)
calculated for the breaking
threshold of the first Cooper pair of nucleons $E_1=0$ MeV
are given in fig. 7.
Its concrete values for minimal $\chi^2$ are equal to 9.6 and 6.3 for
$^{157}$Gd and $^{159}$Gd, respectively. The assumption on energy independence
of $K_{\rm vibr}$ at low excitation energy is evidently unreal
(see fig. 7).

\begin{figure}[htbp]
\vspace{4cm}

\leavevmode
\epsfxsize=17cm

\epsfbox{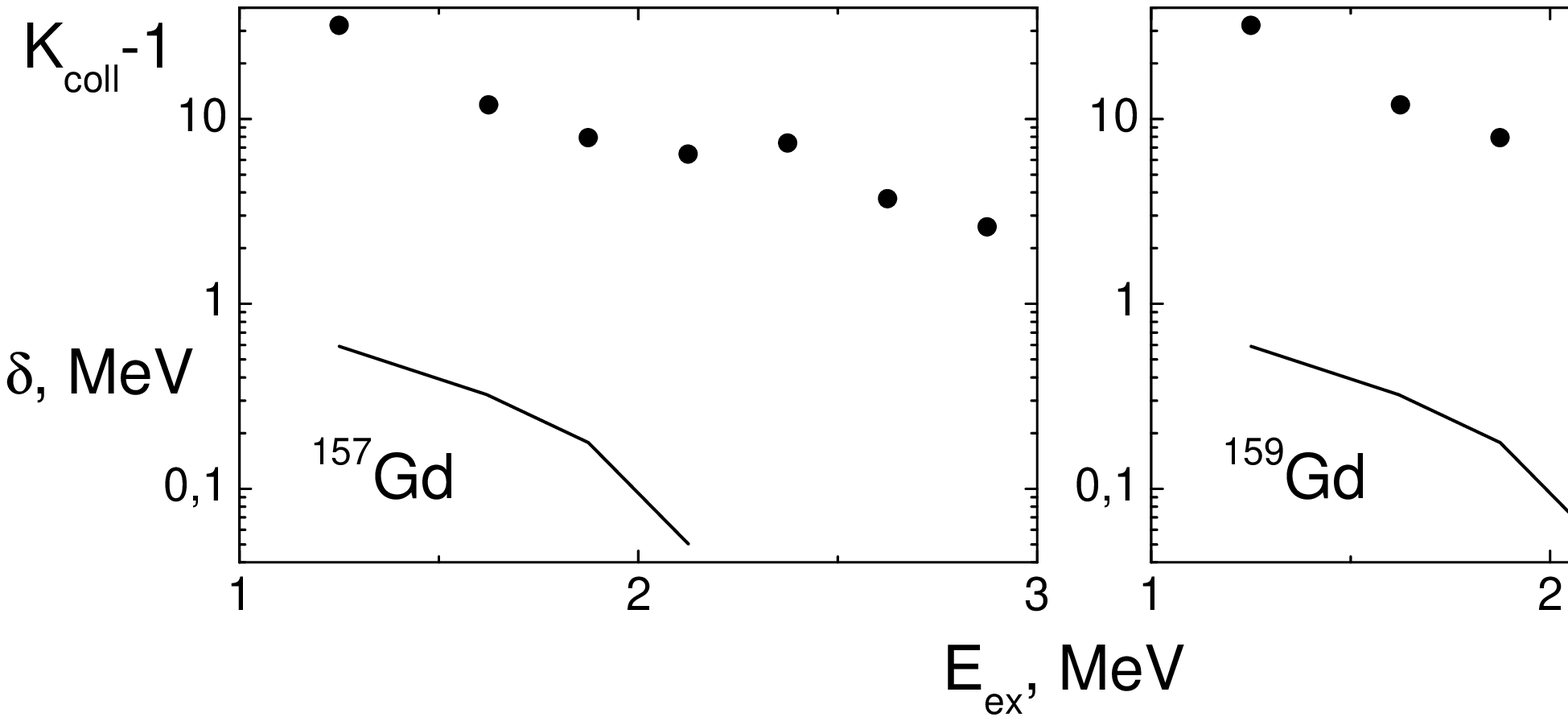}
\vspace{-10.05cm}

{\bf Fig.~7.} Points show coefficient of collective enhancement of
level density, curve represents the values of parameter $\delta_1$
used in  \cite{Prep196,PEPAN-2006} for calculation of partial density of
three-quasiparticle levels.
\end{figure}
Parameter $K_{\rm coll}$-1 determined from comparison between calculated in
this way density of three-quasiparticle excitations ($J=1/2, 3/2$) and its
most probable experimental value is compared with calculated value $\delta_1$
in Fig. 7.

In the excitation energy interval below $\approx 2$ MeV is observed
considerable correlation of this coefficient with the $\delta_1$ value from
\cite{PEPAN-2006} and from the second variant of analysis \cite{Prep196}.
Decrease in correlation at higher excitation energy can be related to
significant contribution of five-quasiparticle excitations in function
$\rho_{qp}(U,J,\pi)$ and/or to smaller than it is adopted in
\cite{Prep196,PEPAN-2006} rate of decrease of function $\delta_1$ at
$U>1.8$ MeV. 

Analysis of experimental data presented above in common with the data
\cite{Meth1,PEPAN-2005} points to necessity of experimental search for
neutron resonance structure in the energy interval $E_n$ of about 1-2 MeV
and more. Figure 6 show both general properties of radiative strength
functions in isotopes under consideration and their evident difference in
region below $\sim 0.5B_n$.

Analogous variations of energy dependence of radiative strength functions
in both nuclei of neighboring elements and isotopes are also demonstrated
by analysis of the two-step cascade intensities. If these changes are
completely or to a great extent determined by difference in ratio between
qusiparticle and phonon components of the neutron resonance wave functions
then extrapolation of the regularity established in \cite{PEPAN-2006} for
expected difference of breaking threshold of two neighboring Cooper pairs
$E_{N+1}-E_N \approx 2\delta$ can be spread and into the region above  $B_n$.
Therefore, one can expect cyclic change in structure of excited resonances
with period of about 2 MeV for heavy nuclei with $\delta\sim 1$ MeV.

Moreover, this effect in even-odd isotopes of rear earth elements can appear
itself as change in ratio between intensities of the primary transitions
with energy $E_1=3-4$ MeV and $E_1 > B_n-1$ MeV.

The data presented allow the following conclusions:

1. Nuclei  $^{157,159}$Gd, excited in the $(\overline{n},\gamma)$ reaction
by the 2 keV and 24 keV neutrons demonstrate the same properties as those
revealed earlier for about forty nuclei from the mass region
$40 \leq A \leq 200$: step-wise structure in level density and local
strengthening of radiative strength functions of the primary gamma-transitions
to the levels of this structure.

2. Abrupt change in structure of levels in the excitation energy region
about 1.0-1.5 MeV. It manifests itself in considerable increase of the
$k(M1)/k(E1)$ values and in strong difference between their distribution
and normal distribution of random gamma-transition amplitudes.

3. These experimental ratios $k(M1)/k(E1)$ can be used for obtaining of simpler
values of the E1- and M1-transition strength functions and data on the ratio
between density of levels with different parity in the frameworks of methods
\cite{Meth1,PEPAN-2005}.

4. Main part of the primary gamma-transitions observed in the
$(\overline{n},\gamma)$ reaction corresponds, probably, to excitation of
levels with large and weakly fragmented phonon components of wave functions.

5. The data on the most probable $N_\gamma$ values obtained for different
intervals of neutron energy allow one to consider practically negligible
dependence of the determined level density on $E_n$. Besides, influence of
nuclear resonance structures results in considerable change in shape of the
deviation distribution of the gamma-transition intensities from its mean
value even in the narrow interval of nuclear excitation energy considered here.

\section{Conclusion}\hspace*{16pt}\hspace*{16pt}

Analysis of the available data on the primary gamma-transition intensities
from $(\overline{n},\gamma)$ reaction in compound nuclei $^{157,159}$Gd
showed step-like structure in density of their levels and increase in
radiative strength function of transitions to  levels in region of this
structure, at least, for  primary dipole gamma-transitions. Id est,
it confirmed main conclusions of \cite{Meth1,PEPAN-2005} and  pointed to
necessity to reveal and remove systematical experimental errors in alternative
methods for determination of only level density \cite{Zh} and simultaneous
determination of all the parameters of cascade gamma-decay \cite{AdNP,NIM}.
The most important problems at experimental determination of level density
and emission probability of reaction product become both correct accounting
for influence of level structure on emission probability of nuclear
evaporation and cascade gamma-quanta in investigations of nuclear reactions
on accelerator beams and considerable decrease in systematical errors of
experiment.

The best estimations of dispersion (parameter $\nu$)	 of the random intensity
fluctuations strongly differ from the mean values predicted in \cite{PT}).
This allows one to assume that the wave functions of levels excited by primary
transitions contain considerable components of weakly fragmented nuclear
states which are more complicated than  one- or three-quasiparticle states.
Approximation of the obtained level density by model \cite{Strut} confirms
the fact of considerable ($\approx$ 10 times) increase of level density due
to excitations of mainly vibration type \cite{PEPAN-2006}.
Comparison of the data presented in figs. 5,6 with those obtained from
intensities of two-step cascades permits one to make preliminary conclusion
that the abrupt change in structure of decaying neutron resonances,
at least, in their energy interval $\approx$ 24 keV is not observed.
There is no reason to wait principle change in the determined according
to method \cite{Meth1,PEPAN-2005} level density and energy dependence
of the primary transition radiative strength functions from resonance
to resonance. Further decrease of errors of these nuclear parameters
determined from intensities of the two-step gamma-cascades undoubtedly
requires reliable estimation of function $k(E_\gamma,E_{ex})$ practically
in all energy diapason of levels excited at thermal neutron capture.


\begin{thebibliography}{99}
\bibitem{Meth1}
	E.~V.~Vasilieva, A.~M. ~Sukhovoj, V.~A.~Khitrov,
	Phys.  At. Nucl., {\bf 64}, 153 (2001).
\bibitem{TSC-err}
	V.~A.~Khitrov,  Li Chol, A.~M.~Sukhovoj, in: 
	 XI International Seminar on Interaction
	of Neutrons with Nuclei,  Dubna, May 2003,
	E3-2004-9, (Dubna, 2004), p. 98;\\
	V.~A.~Khitrov,  Li Chol, A.~M.~Sukhovoj, nucl-ex/0404028.
\bibitem{Appr-TSC}
	 A.~M.~Sukhovoj, V.~A.~Khitrov, Phys. Atomic Nuclei,
	{\bf 62}, 19  (1999).
\bibitem{Yb174}	
	A. M. Sukhovoj,  Physics of Atomic Nuclei, {\bf 71}, 1907 (2008).
\bibitem{Reich}
	  G.W. Reich,Proc. of the 3th International Symposium on
	Capture Gamma-Ray
 	 Spectroscopy and Related Topics, Upton, 18-22 September
 	 1978 /Ed. R.E.Crien, W.R.Kane, Plenum Press, N.Y. and London,p.105.
\bibitem{Malov}
	S.T. Boneva et al., Yad. Fiz.,
	{\bf 49}, 944 (1989).
\bibitem{Gd157} J. Kopecky, M. Uhl, R.E. Chrien, Phys. Rev. 47(1), 312 (1993).
\bibitem{Gd159} G. Granja, S. Pospisil, S.A. Telezhnikov, R.E. Chrien,
	Nucl. Phys. A729,  679 (1984).
\bibitem{MalSol}
	L.A. Malov, V.G. Soloviev, Yad. Fiz.,
	{\bf 26}, 729 (1977).
\bibitem{Stel}
	V.L. Stelts, R.E. Chrien, M.K. Martel, Phys. Rev. C,
	{\bf 24(4)}, 1419 (1981).
\bibitem{PEPAN-2005} A.M. Sukhovoj, V.A. Khitrov, Physics of Particl. and Nuclei,
 	{\bf 36(4)}, 359 (2005).
\bibitem{RIPL}
	Reference Input Parameter Library RIPL-2.  Handbook for
	calculations of nuclear reaction data.  IAEA-TECDOC, 2002.\\
	http://www-nds.iaea.or.at/ripl2.\\ 
	Handbook for Calculation of Nuclear Reactions Data, IAEA, Vienna,
	TECDOC-1034, 1998.
\bibitem{Prep196}
	A.~M.~Sukhovoj, V.~A.~Khitrov,
	Preprint N$^{o}$ E3-2005-196, JINR (Dubna, 2005).
\bibitem{Strut}
	V.~M.~Strutinsky, in  Proc. of Int. Conf. Nucl. Phys.,
	(Paris, 1958), p. 617.
\bibitem{PEPAN-2006}
	A.M. Sukhovoj, V.A. Khitrov, Physics of Paricl. and Nuclei,
	 {\bf 37(6)}, 899 (2006).
\bibitem{Dilg}
	W.~Dilg, W.~Schantl, H.~Vonach, M.~Uhl, Nucl. Phys. A
	{\bf 217}, 269 (1973).	
\bibitem{Axel}
	P.~Axel, Phys. Rev.
	{\bf 126}, 671 (1962).
\bibitem{KMF}
 	S. G. Kadmenskij, V. P. Markushev, V. I. Furman, Sov. J. Nucl. Phys.,
 	{\bf 37}, 165 (1983).
\bibitem{Zh} O.T.  Grudzevich et.al., Sov.  J.  Nucl.  Phys. {\bf 53}, 92 (1991).\\
 B.V.  Zhuravlev, Bull. Rus. Acad. Sci. Phys. {\bf 63} 123 (1999).
\bibitem{AdNP}
	G.~A.~Bartholomew  et al., Advances in nuclear physics
	{\bf 7}, 229 (1973).
\bibitem{NIM}
	 A.~Schiller {\it et al.}, Nucl.  Instrum. Methods A
	{\bf 447}, 498 (2000).
\bibitem{PT}
	C.~F.~Porter, R.~G.~Thomas, Phys. Rev.
	{\bf 104}, 483 (1956).
\end{thebibliography}
\end{document}